\documentclass[aps,prb,twocolumn,superscriptaddress,showkeys]{revtex4}

\usepackage{graphicx}

\begin{document}

\title{Influence of Fe and Co on Phase Transitions in Ni-Mn-Ga Alloys}

\author{Vladimir V. Khovailo}
\author{Toshihiko Abe}
\affiliation{National Institute of Advanced Industrial Science and
Technology, Tohoku Center, Sendai 983-8551, Japan}

\author{Viktor V. Koledov}
\affiliation{Institute of Radioengineering and Electronics of RAS,
Moscow 125009, Russia}

\author{Minoru Matsumoto}
\affiliation{Institute of Multidisciplinary Research for Advanced
Materials, Tohoku University, Sendai 980-8577, Japan}

\author{Hirokazu Nakamura}
\affiliation{Institute of Fluid Science, Tohoku University, Sendai
980-8577, Japan}

\author{Ryunosuke Note}
\affiliation{Institute for Material Research, Tohoku University,
Sendai 980-8577, Japan}

\author{Makoto Ohtsuka}
\affiliation{Institute of Multidisciplinary Research for Advanced
Materials, Tohoku University, Sendai 980-8577, Japan}

\author{Vladimir G. Shavrov}
\affiliation{Institute of Radioengineering and Electronics of RAS,
Moscow 125009, Russia}

\author{Toshiyuki Takagi}
\affiliation{Institute of Fluid Science, Tohoku University, Sendai
980-8577, Japan}

\begin{abstract}
Differential scanning calorimetry (DSC) and magnetic measurements
were performed to study the influence of ferromagnetic 3-$d$
transition elements Fe and Co on structural and magnetic
properties of ferromagnetic shape memory alloys Ni$_2$MnGa.
Addition of Fe or Co on the Ni sites decreases the temperature of
martensitic phase transition $T_m$, whereas addition of Co on the
Mn sites results in a considerable increase of $T_m$. Magnetic
measurement revealed that Curie temperature $T_C$ increases upon
substitution of Fe or Co for Ni. This observation is of importance
for design of high temperature ferromagnetic shape memory alloys.
\end{abstract}


\maketitle

\section{Introduction}

In resent years ferromagnetic shape memory alloys have attracted
considerable attention as a new class of actuator materials (see,
for a recent review, Ref.~1). Among ferromagnetic shape memory
alloys, the largest magnetic-field-induced strains arising from
conversion of martensitic variants under action of a magnetic
field~\cite{2-m,3-u} or caused by the shift of the martensitic
transition temperature~\cite{4-c,5-t} have been observed in the
most intensively studied Ni$_2$MnGa alloy system.

Ni$_2$MnGa is a representative of Mn-containing Heusler alloys.
For the stoichiometric composition a structural phase transition
of martensitic type from the parent cubic to a complex
tetragonally based structure occurs at $T_m = 202$~K, whereas
long-range ferromagnetic ordering sets at $T_C =
376$~K~\cite{6-w}. The martensitic transition temperature $T_m$
was found to be very sensitive to the composition, ranging from
liquid helium temperature up to over 600~K~\cite{7-c,8-j}.
Contrary to $T_m$, Curie temperature $T_C$ of Ni-Mn-Ga alloys is
less composition dependent. Based on the published experimental
results~\cite{9-w} it can be concluded that the highest $T_C
\approx 380$~K is observed in the stoichiometric Ni$_2$MnGa. A
decrease of $T_C$ in the alloys with deficiency in Mn, which
possesses a magnetic moment of $\sim 4~\mu _B$, is due to the
dilution of the magnetic subsystem. For the alloys with Mn excess,
it was suggested that the decrease in $T_C$ is accounted for by
antiferromagnetic coupling of the extra Mn atoms~\cite{10-e}.

Analyzing data collected for a broad compositional range of
Ni-Mn-Ga alloys, Chernenko~\cite{11-c} pointed out that there
exists a correlation between the electron concentration and
stability of high-temperature $\beta$-phase in the Ni-Mn-Ga
system. Now it is generally acknowledged that a change of electron
concentration $e/a$ significantly affects martensitic transition
temperature $T_m$ of Ni-Mn-Ga alloys, and, therefore, an empirical
relationship between $T_m$ and $e/a$ can be reasonably used for
the preparation of alloys with required martensitic transition
temperature~\cite{12-j}. Since large magnetic-field-induced
strains are observed only in the ferromagnetic state, an efficient
control of the ferromagnetic ordering temperature is of importance
for realization of this effect in a large temperature interval and
development of high temperature ferromagnetic shape memory alloys.
Since previous experimental studies~\cite{9-w,11-c} have shown
that any deviation from the stoichiometry results in decrease of
$T_C$ in Ni-Mn-Ga alloys, we studied the influence of 3-$d$
ferromagnetic transition elements Fe and Co on Curie temperature
$T_C$ and martensitic transition temperature $T_m$ of Ni-Mn-Ga.

\section{Sample preparation and measurements}

Co- and Fe-containing Ni-Mn-Ga ingots of various compositions were
prepared by arc-melting of high-purity initial elements in argon
atmosphere. The ingots were annealed at 1100~K for 9~days in
quartz ampoules and quenched in ice water. Samples for the
measurements were cut from the middle part of the ingots. The real
compositions of the alloys were determined by an Inductively
Coupled Plasma Mass Spectrometry (ICPMS) method. Determination of
the chemical compositions by ICPMS revealed that the real
compositions of the alloys are close to the nominal ones.
Characteristic temperatures of the direct (martensite start, $M_s$
and martensite finish, $M_f$) and the reverse (austenite start,
$A_s$ and austenite finish, $A_f$) martensitic transformations
were determined from differential scanning calorimetry (DSC)
measurements. The martensitic transition temperature $T_m$ was
calculated as $T_m = (M_s + A_s)/2$. A vibrating sample
magnetometer (VSM) was used for low-field magnetic measurements.
Curie temperature $T_C$ was determined as a minimum on the
temperature derivative of magnetization curve measured upon
heating, $dM/dT$. Nominal compositions of the alloys studied
together with their martensitic and ferromagnetic transition
temperatures are given in Table~1.

\begin{table}
\caption{Nominal composition, martensitic transition temperature
$T_m$ and Curie temperature $T_C$ of the alloys studied.}
\begin{tabular}{|c|c|c|}

\hline

Nominal composition  &  $T_m$ (K) & $T_C$ (K) \\

\hline

Ni$_{2.13}$Co$_{0.03}$Mn$_{0.84}$Ga & 318 & 348 \\

Ni$_{2.10}$Co$_{0.06}$Mn$_{0.84}$Ga & 315 & 361 \\

Ni$_{2.07}$Co$_{0.09}$Mn$_{0.84}$Ga & 306 & 367 \\

Ni$_{2.16}$Mn$_{0.80}$Co$_{0.04}$Ga & 345 & 352 \\

Ni$_{2.16}$Mn$_{0.75}$Co$_{0.09}$Ga & 437 & 372 \\

Ni$_{2.16}$Mn$_{0.70}$Co$_{0.14}$Ga & 525 & 350 \\

Ni$_{2.16}$Fe$_{0.04}$Mn$_{0.80}$Ga & 317 & 347 \\

Ni$_{2.12}$Fe$_{0.08}$Mn$_{0.80}$Ga & 289 & 358 \\

Ni$_{2.08}$Fe$_{0.12}$Mn$_{0.80}$Ga & 267 & 377 \\

Ni$_{2.04}$Fe$_{0.16}$Mn$_{0.80}$Ga & 223 & 395 \\

 \hline

\end{tabular}
\end{table}

\section{Experimental results}

\subsection{Co-containing alloys}

The nominal compositions of the alloys studied are
Ni$_{2.16-x}$Co$_x$Mn$_{0.84}$Ga ($x$ = 0.03, 0.06 and 0.09) and
Ni$_{2.16}$Mn$_{0.84-y}$Co$_y$Ga ($y$ = 0.04, 0.09 and 0.14). It
was found that the compositional dependencies of $T_m$ strongly
depend on the Co position. Thus, the martensitic transition
temperature $T_m$ of the Ni$_{2.16-x}$Co$_x$Mn$_{0.84}$Ga alloys
slightly decreases with the Co excess, whereas in the
Ni$_{2.16}$Mn$_{0.84-y}$Co$_y$Ga alloys the increase in Co content
results in a significant increase of $T_m$ (Table~1). A marked
increase in the temperature of martensitic transition in the
alloys where Co is substituted for Mn makes these materials
attractive for use as high-temperature shape memory alloys.

\begin{figure}[h]
\begin{center}
\includegraphics[width=7cm]{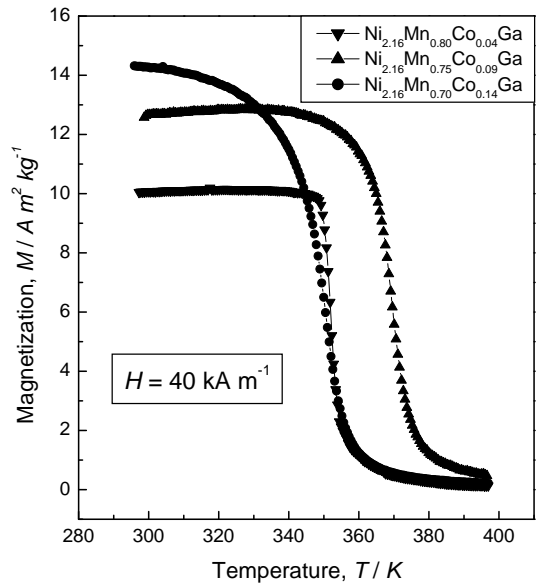}
\caption{Temperature dependencies of magnetization $M$ for the
Ni$_{2.16}$Mn$_{0.84-y}$Co$_y$Ga alloys.}
\end{center}
\end{figure}

The behavior of Curie temperature $T_C$ in these two series of
alloys also depends significantly on the position of Co. In the
Ni$_{2.16-x}$Co$_x$Mn$_{0.84}$Ga alloys Curie temperature
increases from 348~K to 367~K with increasing Co content from $x =
0.03$ to $x = 0.06$ (Table~1). Contrary to this, $T_C$ in the
Ni$_{2.16}$Mn$_{0.84-y}$Co$_y$Ga alloys exhibits a non-monotonic
dependence, and initially increases from $T_C = 352$~K ($y =
0.04$) to $T_C = 372$~K ($y = 0.09$) and then decreases to $T_C =
350$~K for $y = 0.14$ (Fig.~1). Moreover, $T_C$ in the
Ni$_{2.16}$Mn$_{0.80}$Co$_{0.04}$Ga alloy exhibits hysteretic
feature, as evidenced from $M(T)$ measured upon heating and
cooling (Fig.~2). Such an unusual behavior of ferromagnetic
transition temperature is caused by the coupling of martensitic
and ferromagnetic transition temperatures and has been observed in
Ni$_{2.18}$Mn$_{0.82}$Ga and Ni$_{2.19}$Mn$_{0.81}$Ga
alloys~\cite{13-k,14-f}.

\begin{figure}[h]
\begin{center}
\includegraphics[width=7cm]{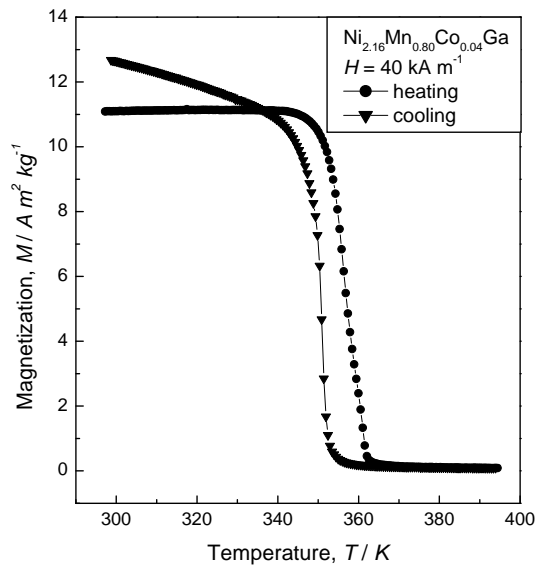}
\caption{$M(T)$ measured upon heating and cooling for the
Ni$_{2.16}$Mn$_{0.80}$Co$_{0.04}$Ga alloy.}
\end{center}
\end{figure}

\subsection{Fe-containing alloys}

In the Fe-containing series of Ni-Mn-Ga alloys Fe was added
instead of Ni to the host Ni$_{2.20}$Mn$_{0.80}$Ga composition.
The nominal compositions of Ni$_{2.20-z}$Fe$_z$Mn$_{0.80}$Ga
alloys are characterized by the Fe content $z$ = 0.04, 0.08, 0.12
and 0.16. DSC measurements revealed that the martensitic
transition temperature continuously decreases from $T_m = 317$~K
to $T_m = 223$~K as Fe content increases from $z = 0.04$ to $z =
0.16$ (Table~1).

The low-field magnetization measurements showed that Curie
temperature $T_C$ of the Ni$_{2.20-z}$Fe$_z$Mn$_{0.80}$Ga alloys
increases with the increase in Fe content (Table~1). In the sample
with the highest Fe content, Ni$_{2.04}$Fe$_{0.16}$Mn$_{0.80}$Ga,
$T_C = 395$~K is higher than that of the stoichiometric
Ni$_2$MnGa.

\section{Discussion}

It was suggested that the electron concentration plays an
important role not only in stabilizing the Heusler structure but
that it is also a driving force for the structural phase
transition in Ni$_2$MnGa, which takes place when the contact
between the Fermi surface and Brillouin zone boundary
occurs~\cite{6-w}. The change in a number of valence electrons
results in alteration of the density of states curve, which is a
basis for the correlation the extent of phase stability and
electron concentration $e/a$~\cite{15-s}.

Our study of Co- and Fe-containing Ni-Mn-Ga alloys revealed that
increasing/decreasing in the number of conduction electron results
in increase/decrease of the martensitic transformation
temperature. Data presented in Fig.~3 clearly demonstrate this
tendency. Resulting in decrease of the electron concentration
$e/a$, substitution of Co for Ni as well as Fe for Ni leads to the
decrease of martensitic transition temperature. In the case of
substitution of Co for Mn, the increase in the electron
concentration brings to the increase of $T_m$ from 345~K for
Ni$_{2.16}$Mn$_{0.80}$Co$_{0.04}$Ga to 518~K for
Ni$_{2.16}$Mn$_{0.70}$Co$_{0.14}$Ga. This result is in agreement
with an earlier experimental study of Co-containing Ni-Mn-Ga
alloys~\cite{16-n}.

\begin{figure}[h]
\begin{center}
\includegraphics[width=7cm]{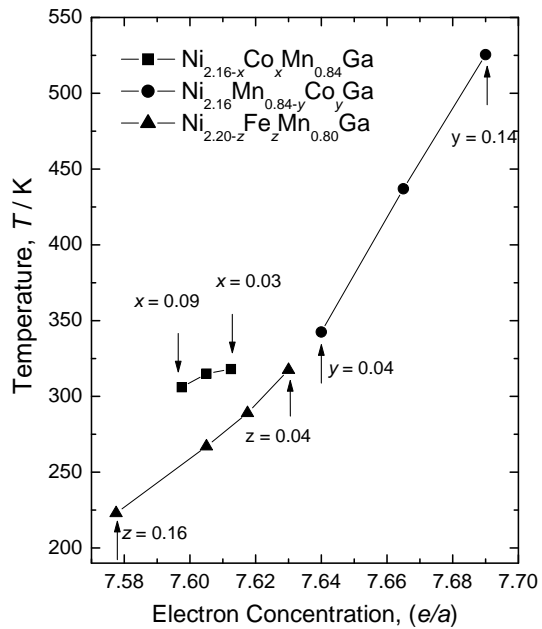}
\caption{Martensitic transition temperature $T_m$ of the Fe- and
Co-containing Ni-Mn-Ga alloys as a function of electron
concentration $e/a$.}
\end{center}
\end{figure}

Therefore, these results indicate that the approach used for the
description of martensitic transition temperature as a function of
electron concentration in ternary Ni-Mn-Ga can also be applied for
the case of quaternary Heusler alloys. Obviously, such an approach
can be used only in a limited range of concentration of the forth
element, where phase separation does not take place. It is worth
noting, however, that new ferromagnetic shape memory alloys were
found recently in Ni-Co-Ga~\cite{17-o} and
Ni-Fe-Ga~\cite{18-o,19-o,20-l,21-l} systems, which formally can be
considered as the case of complete substitution of Mn for Co and
Fe, respectively.

Magnetization measurements of two series of samples where Ni was
substituted by Co or Fe showed that Curie temperature increases
with increasing Co (or Fe) content (Table~1). The most pronounced
increase of $T_C$ is observed in the Fe-containing alloys, where
Curie temperature increases from 347~K ($z = 0.04$) to 395~K ($z =
0.16$). Note that a considerable increase in $T_C$ was also
observed in alloys where Fe was added instead of Mn,
Ni$_2$Mn$_{1-x}$Fe$_x$Ga~\cite{22-l,23-w}. The highest Curie
temperature, $T_C = 433$~K, reported in those studies, was
observed in the Ni$_{2.016}$Mn$_{0.40}$Fe$_{0.72}$Ga$_{0.864}$
composition.

The Mn content remains constant in the Fe- and Co-containing
alloys (Mn content $y = 0.84$ for the Co-containing samples and Mn
content $z = 0.80$ for the Fe-containing samples) and, therefore,
the Mn magnetic subsystem is supposed to be not influenced by
these substitutions. The observed increase of $T_C$ in the Co- and
Fe-containing samples implies that the magnetic properties of
Ni-Mn-Ga should be considered taking into account small magnetic
moments located on Ni atoms~\cite{24-w} and their coupling with
magnetic moments of Mn atoms. For example, the increase in $T_C$
in these alloys can be accounted for by a stronger Co-Mn (Fe-Mn)
exchange interaction as compared to the Ni-Mn one. Our
experimental data, however, are not sufficient to make an
unambiguous conclusion about the mechanism responsible for the
increase of $T_C$ and therefore this effect needs further
investigation. The experimentally observed increase of $T_C$ in
Ni$_{2.16-x}$Co$_x$Mn$_{0.84}$Ga and
Ni$_{2.20-z}$Fe$_z$Mn$_{0.80}$Ga is in accordance with a recent
theoretical consideration of magnetic properties of Ni$_2$MnGa and
Ni$_2$MnAl~\cite{25-e}. In Ref. [25] a suggestion was made that Ni
significantly affects $T_C$ in these alloys and, therefore, Curie
temperature in Ni$_2$MnGa could be increased by a substitution of
the Ni sites. Note also that this suggestion is in accordance with
empirical observation that Curie temperature of Co$_2$MnZ or
Cu$_2$MnZ (Z is a non-transition element) alloys is higher than
$T_C$ in corresponding Ni$_2$MnZ alloys~\cite{24-w}.

In the Ni$_{2.16}$Mn$_{0.84-y}$Co$_y$Ga alloys Curie temperature
$T_C$ exhibits a non-monotonous dependence as the Co content
increases (Fig.~1). This is rather unusual, because in the
Mn-containing Heusler alloys the dilution of the Mn magnetic
subsystem generally results in decrease of $T_C$. In
Ni$_{2+x}$Mn$_{1-x}$Ga alloys, however, merging of $T_m$ and $T_C$
is accompanied by the increase of $T_C$ in a limited composition
interval $x = 0.18 - 0.20$~\cite{26-v}. By analogy with the
Ni$_{2+x}$Mn$_{1-x}$Ga alloys it could be suggested that the
increase of $T_C$ from 352~K to 372~K as the Co content changes
from $y = 0.04$ to $y = 0.09$ is accounted for by the coupling of
martensitic and ferromagnetic transitions, occurring in the
Ni$_{2.16}$Mn$_{0.75}$Co$_{0.09}$Ga alloy. This is not the case,
as evident from the DSC measurements, which revealed that for this
composition $T_m = 437$~K is considerably higher than $T_C$.
Therefore, it is necessary to look for another mechanism
responsible for the non-monotonous compositional dependence of
$T_C$ in the Ni$_{2.16}$Mn$_{0.84-y}$Co$_y$Ga alloys. It is worth
noting that similar behavior of $T_C$ has been observed in a
Fe$_2$Mn$_{1-x}$V$_x$Si system, where Curie temperature was found
to increase from 219~K to 315~K with increasing $x$ from $x = 0$
to $x = 0.5$, followed by a rapid decrease at a higher V
content~\cite{27-k}. It was suggested that this unusual behavior
of the ferromagnetic ordering temperature in
Fe$_2$Mn$_{1-x}$V$_x$Ga could be related to the band structure of
the system~\cite{28-f}. In future it would be interesting to
perform similar band calculation for the
Ni$_{2.16}$Mn$_{0.84-y}$Co$_y$Ga system. Further experimental
studies of Ni$_{2.16}$Mn$_{0.84-y}$Co$_y$Ga alloys are also
necessary in order to determine the critical Co content at which
Curie temperature has a maximal value.

\section{Conclusion}

In conclusion, our experimental results have revealed that in the
studied Fe- and Co-containing Ni-Mn-Ga alloys the behavior of the
martensitic transition temperature can be satisfactory described
as a function of the electron concentration $e/a$. The magnetic
measurements have shown that addition of Fe or Co instead of Ni
results in increase of Curie temperature $T_C$. This observation
together with the theoretical suggestion~\cite{25-e} gives a basic
ground for the development of high temperature ferromagnetic shape
memory alloys in the Ni-Mn-Ga system.



\end{document}